\documentclass[12pt]{article}
\pdfoutput=1
\usepackage{graphicx,amsfonts,amsthm,amsmath,amssymb,hyperref,color,yfonts,cite}
\newcommand{\Tr}{{\rm Tr}}
\newcommand{\bea}{\begin{eqnarray}\displaystyle}
\newcommand{\eea}{\end{eqnarray}}

\begin{document}
\makeatletter
\@addtoreset{equation}{section}
\makeatother
\renewcommand{\theequation}{\thesection.\arabic{equation}}
\vspace{1.8truecm}

{\LARGE{ \centerline{\bf Constructive Holography}}}  

\vskip.5cm 

\thispagestyle{empty} 
\centerline{{\large\bf Robert de Mello Koch$^{a,b,d}$\footnote{{\tt robert@zjhu.edu.cn}} and
 Hendrik J.R. Van Zyl a$^{c,d}$\footnote{\tt hjrvanzyl@gmail.com}}}
\vspace{.8cm}
\centerline{{\it $^{a}$School of Science, Huzhou University, Huzhou 313000, China,}}
\vspace{.8cm}
\centerline{{\it $^{b}$School of Physics and Mandelstam Institute for Theoretical Physics,}}
\centerline{{\it University of the Witwatersrand, Wits, 2050, }}
\centerline{{\it South Africa }}
\vspace{.8cm}
\centerline{{\it $^{c}$The Laboratory for Quantum Gravity \& Strings,}}
\centerline{{\it Department of Mathematics \& Applied Mathematics,}}
\centerline{{\it University of Cape Town, Cape Town, South Africa}}
\vspace{.8cm}
\centerline{{\it $^{d}$ The National Institute for Theoretical and Computational Sciences,}} \centerline{{\it Private Bag X1, Matieland, South Africa}}

\vspace{1truecm}

\thispagestyle{empty}

\centerline{\bf ABSTRACT}

\vskip.2cm 
We consider the collective field theory description of the singlet sector of a free and massless matrix field in $d$ dimensions. The $k$-local collective fields are functions of $(d-1)k+1$ coordinates. We provide a map between the collective fields and fields in the dual gravitational theory defined on AdS$_{d+1}$ spacetime. The coordinates of the collective field have a natural interpretation: the $k$-local collective field is a field defined on an AdS$_{d+1}\times$S$^{k-1}\times$S$^{(d-2)(k-2)}\times$S$^{d-3}$ spacetime. The modes of a harmonic expansion on the S$^{k-1}\times$S$^{(d-2)(k-2)}\times$S$^{d-3}$ portion of the spacetime leads to the spinning bulk fields of the dual gravity theory.

\setcounter{page}{0}
\setcounter{tocdepth}{2}
\newpage
\tableofcontents
\setcounter{footnote}{0}
\linespread{1.1}
\parskip 4pt

{}~
{}~

\section{Introduction}

An important clue into the AdS/CFT duality \cite{Maldacena:1997re,Gubser:1998bc,Witten:1998qj} is the observation that the loop expansion parameter of the conformal field theory ($\hbar$) is replaced by ${1\over N}$. Collective field theory \cite{Jevicki:1979mb,Jevicki:1980zg}, which systematically reorganizes the degrees of freedom within the field theory, results in a theory with a loop expansion parameter of ${1\over N}$. This makes it a natural candidate for a constructive approach to the duality. Within the collective field theory framework, the field theory dynamics are expressed in terms of an over complete set of invariant field variables known as collective fields. When interpreted as a constructive approach to the AdS/CFT correspondence, the theory of collective fields is to be identified with the dual gravitational dynamics. The overcompleteness of the collective fields has been suggested as the underlying reason for the holographic nature of the collective description\cite{deMelloKoch:2023ylr}.

Vector models provide a straightforward framework for the detailed development of collective field theory. The pioneering work \cite{Das:2003vw} first suggested the relevance of collective field theory for holography within the context of a free vector model in three dimensions. The gravitational dual in this case is a theory of higher spin gravity \cite{Klebanov:2002ja,Sezgin:2002rt,Vasiliev:1990en}. In this framework, the collective fields are bilocal fields dependent on five coordinates, four of which describe the dual AdS bulk. The proposal of \cite{Das:2003vw} is that the extra coordinate serves to repackage the entire tower of fields in the gravitational theory into a single field. This proposal was realised in detail in \cite{deMelloKoch:2010wdf}, utilizing light front quantization. This initial work was extended to a standard equal time description in \cite{deMelloKoch:2014vnt}, to finite temperature in \cite{Jevicki:2015sla,Jevicki:2017zay,Jevicki:2021ddf}, to $d$ dimensions in \cite{Jevicki:2011ss,Jevicki:2011aa}, to the IR fixed point in \cite{Mulokwe:2018czu,Johnson:2022cbe} and to a covariant two time description in \cite{deMelloKoch:2018ivk,Aharony:2020omh}. The collective field description reproduces the correct equations of motion and boundary conditions \cite{deMelloKoch:2010wdf,deMelloKoch:2014vnt,Mintun:2014gua} and so constitutes a complete bulk reconstruction. In addition, information localizes \cite{deMelloKoch:2021cni,deMelloKoch:2022sul,deMelloKoch:2023ngh,deMelloKoch:2023ylr,deMelloKoch:2024juz} exactly as expected in a theory of quantum gravity \cite{Laddha:2020kvp}. Taken together, this is compelling evidence that collective field theory does indeed provide a constructive approach to holography.

Recently \cite{deMelloKoch:2024ewt,deMelloKoch:2024otg} the collective field theory description of free matrix models in $d=3$ dimensions was considered.  The gravitational theory dual to the matrix conformal field theory (CFT) is a tensionless string theory defined on AdS$_{d+1}$ \cite{Mikhailov:2002bp}. The paper \cite{deMelloKoch:2024ewt} established a match between the conformal symmetry generators of the collective field theory and those of the dual gravity theory, as well as between the equations of motion in the dual gravity and those in the collective field theory. This study make good use of the light cone gauge description of spinning massless and massive fields on the bulk AdS$_{d+1}$ spacetime which has been developed in an impressive series of papers \cite{Metsaev:1999ui,Metsaev:2003cu,Metsaev:2005ws,Metsaev:2013kaa,Metsaev:2015rda,Metsaev:2022ndg}. The paper \cite{deMelloKoch:2024otg} was concerned with the interpretation of the extra coordinates parametrizing the collective field. To get some insight into why this interpretation is needed, consider a collective field theory treatment of a free massless matrix in $d$ dimensions. This conformal field theory is described by the action
\bea
S&=&\int d^d x\,{1\over 2}\,\Tr \big(\partial_\mu\phi\,\partial^\mu\phi\big)
\eea
where $\phi$ is an $N\times N$ matrix valued field. The corresponding Hamiltonian is
\bea
H={1\over 2}\int d^{d-1}x\,\Tr\, \left(\pi(t,\vec{x})\,\pi(t,\vec{x})+\vec{\nabla}\phi\, \cdot\,\vec{\nabla}\phi\right)
\eea
where the conjugate momentum is
\bea
\pi(t,\vec{x})_{ab}&=&{1\over i}{\delta\over\delta\phi (t,\vec{x})_{ba}}
\eea
A suitable set of collective fields are given by the $k$-local equal time invariants
\bea
\sigma_k(t,\vec{x}_1,\cdots,\vec{x}_k)&\equiv&\Tr\,\left(\phi(t,\vec{x}_1)\phi(t,\vec{x}_2)\cdots \phi(t,\vec{x}_k)\right) 
\eea
We can rewrite the Hamiltonian in terms of the momentum conjugate to $\sigma_k(t,\vec{x}_1,\cdots,\vec{x}_k)$ denoted $\Pi_k(t,\vec{x}_1,\cdots,\vec{x}_k)$ and given by
\bea
\Pi_k(t,\vec{x}_1,\cdots,\vec{x}_k)&=&{1\over i}{\delta\over\delta \sigma_k(t,\vec{x}_1,\cdots,\vec{x}_k)}
\eea
by using the functional chain rule as explained in the original paper \cite{Jevicki:1979mb}. The resulting Hamiltonian is no longer manifestly Hermitian since the change of variables is necessarily accompanied by a non-trivial measure given by a Jacobian. If one transforms to a description with trivial measure, the requirement that the Hamiltonian is manifestly Hermitian then determines the Jacobian \cite{Jevicki:1979mb}. This Jacobian is a highly non-linear function of the collective fields and it determines an infinite sequence of interaction vertices. This non-linear interacting theory is the collective field description.

It is significant that the collective field and its conjugate momentum obey the usual equal time commutation relations
\bea
[\Pi_k(t,\vec{x}_1,\cdots,\vec{x}_k),\sigma_k(t,\vec{y}_1,\cdots,\vec{y}_k)]&=&-i\prod_{i=1}^k \delta^{(d-1)}(\vec{x}_i-\vec{y}_i)
\eea
This, implies, as usual, that we have quantized a degree of freedom living at each distinct $(\vec{x}_1,\vec{x}_2,\cdots,\vec{x}_k)$ independently. Thus, if collective field theory does indeed provide a construction of the dual gravitational theory, the complete set of $k(d-1)$ spatial coordinates of the $k$-local collective field must have a sensible interpretation as coordinates of some space in the dual gravity description. Finding this interpretation for the collective field theory of the $d=3$ free matrix conformal field theory is the key result achieved in \cite{deMelloKoch:2024otg}. This paper shows that the $k$-local collective field lives on an AdS$_4\times$S$^{k-2}\times$S$^{k-1}$ spacetime. When taking the operator product of $k$ matrix fields many primary operators are generated, each of which corresponds to a field in the dual gravity. The extra coordinates, which parametrize S$^{k-2}\times$S$^{k-1}$, collect this complete set of gravitational fields into a single field. The fields of the dual gravity theory can be recovered by performing a harmonic expansion on the S$^{k-2}\times$S$^{k-1}$ space.

The goal of this paper is to extend the results of \cite{deMelloKoch:2024ewt,deMelloKoch:2024otg} to general $d\ge 3$ dimensions. Following \cite{deMelloKoch:2010wdf}, we work with equal $x^+$ collective fields 
\bea
\sigma_k(x^+,x_1^-,\vec{x}_1,x_2^-,\vec{x}_2,\cdots,x_k^-,\vec{x}_k)&=&{\rm Tr}\big(\phi(x^+,x_1^-,\vec{x}_1)\phi(x^+,x_2^-,\vec{x}_2)\cdots\phi(x^+,x_k^-,\vec{x}_k)\big)\cr
&&
\eea
where $\vec{x}_i$ is a vector collecting coordinates transverse to the light cone, $x^+$ is the shared light cone time and $x_i^-$ is the second light cone coordinate. It is clear that the $k$-local collective fields are functions of $(d-1)k+1$ coordinates. The holographic mapping between the conformal field theory and the tensionless string theory involves an identification between the fields of the two theories, as well as an identification between the coordinates. Following \cite{deMelloKoch:2010wdf}, the map is determined by requiring that it takes the realization of conformal symmetry on the collective fields into that of the dual gravity fields. Concretely this amounts to a matching of the generators of conformal symmetry in the collective field theory with those of the tensionless string. We carry out the matching of conformal symmetry in Section \ref{matchconf}, which determines the holographic map. We consider the interpretation of the extra bulk coordinates in Section \ref{Bcoords}. We find that these coordinates parametrize a product of spheres S$^{k-1}\times$S$^{(d-2)(k-2)}\times$S$^{d-3}$. By studying the behaviour of the reconstructed bulk fields near the boundary ($Z\to 0$) in Section \ref{BC} we argue that the modes of a harmonic expansion on the S$^{k-1}\times$S$^{(d-2)(k-2)}\times$S$^{d-3}$ portion of the spacetime leads to the spinning bulk fields of the dual gravity theory. In Section \ref{conclusions} we wrap up the paper with conclusions and suggestions about how this study might be extended.

\section{Matching Conformal Symmetries}\label{matchconf}

The mapping between collective field theory and gravity involves two key components: the relationship between the fields and the relationship between the coordinates of the two theories. Both components are determined by matching the conformal symmetry of the boundary and bulk theories.

The mapping between the fields involves a similarity transformation. By rewriting the conformal symmetry generators of the collective field theory using the coordinate map and applying the similarity transformation, the generators of the gravity theory are exactly reproduced. Thus, to determine the mapping between bulk and boundary theories, we use the conformal symmetry generators from both descriptions as input. The generators of the collective field theory are derived from the coproduct of the usual generators for the free scalar field in $d$ dimensions. The bulk AdS$_{d+1}$ generators have been established in a series of remarkable papers \cite{Metsaev:1999ui, Metsaev:2003cu, Metsaev:2005ws, Metsaev:2013kaa, Metsaev:2015rda, Metsaev:2022ndg}.

To specify the mapping between fields, recall that the collective field acquires a large $N$ expectation value. By expanding around this large $N$ background, we define fluctuations $\eta_k(x^+,x_1^-,\vec{x}_1,\cdots,x_k^-,\vec{x}_k)$ as follows
\bea
\sigma_k(x^+,x_1^-,\vec{x}_1,\cdots,x_k^-,\vec{x}_k)&=&\sigma_k^0(x^+,x_1^-,\vec{x}_1,\cdots,x_k^-,\vec{x}_k)+{1\over N^{k\over 2}}\eta_k(x^+,x_1^-,\vec{x}_1,\cdots,x_k^-,\vec{x}_k)\cr
&&
\eea
For odd $k$ $\sigma_k^0(x^+,x_1^-,\vec{x}_1,\cdots,x_k^-,\vec{x}_k)$ vanishes while for even $k$ it is of order $N^{{k\over 2}+1}$. The coefficient of the fluctuation is chosen so that the two point function of $\eta_k$ is order 1 as $N\to\infty$. The holographic mapping identifies the fluctuation $\eta_k$ with a bulk gravity field $\Phi$. The analysis is most easily carried out \cite{deMelloKoch:2021cni} after performing a Fourier transform in the bulk\footnote{We will consistently use capital letters for the coordinates of the AdS bulk spacetime and little letters for the coordinates of the collective field theory.} which replaces $X^-$ with $P^+$, and a corresponding Fourier transform in the boundary which replaces $x^-$ with $p^+$. The relation between the fields of the collective field theory and those of the bulk gravity takes the form
\bea
\Phi(X^+,P^+,\vec{X},Z,\{\alpha_i\})
=\mu(\vec{x}_i,p_i^+)\eta_k(x^+,p_1^+,\vec{x}_1,\cdots,p_k^+,\vec{x}_k)\label{relflds}
\eea
The extra coordinates $\{\alpha_i\}$, as well as their role, will be discussed in detail in Section \ref{Bcoords}. That analysis will establish that the expansion coefficients obtained by expanding the reconstructed bulk field $\Phi(X^+,P^+,\vec{X},Z,\{\alpha_i\})$ in a complete set of orthogonal functions of the $\{\alpha_i\}$ define the fields of the dual AdS$_{d+1}$ gravity theory.

The generators acting on the $k$-local collective field are obtained by taking the coproduct of the generators of the usual free scalar field. The generators are\footnote{Letters from the begining of the alphabet, $a,b,...$ run over the directions transverse to the lightcone ($1,2,...,d-2$) while letters from the middle of the alphabet $i,j,...$ run over the different fields appearing in the $k$-local collective field ($1,2,...,k$).}
\begin{eqnarray}
P^+&=&\sum_{i=1}^k p_i^+\cr\cr
\vec{P}&=&\sum_{i=1}^k\,{\partial\over \partial \vec{x}_i}\cr\cr
P^-&=&-\sum_{i=1}^k\, {1\over 2p_i^+}{\partial\over\partial \vec{x}_i}\cdot{\partial\over\partial \vec{x}_i}\cr\cr
J^{+-}&=&x^+ P^- +\sum_{i=1}^k\, {\partial\over\partial p_i^+}\, p_i^+\cr\cr
J^{+a}&=&x^+\sum_{i=1}^k\,{\partial\over\partial x^a_i}-\sum_{i=1}^k\,x^a_i p_i^+\cr\cr
J^{-a}&=&\sum_{i=1}^k\,\left(-{\partial\over\partial p_i^+}{\partial\over\partial x^a_i}+{x^a_i\over 2p_i^+}{\partial\over\partial \vec{x}_i}\cdot{\partial\over\partial \vec{x}_i}\right)\cr\cr
J^{ab}&=&\sum_{i=1}^k \left(x^a_i{\partial\over\partial x^b_i}-x^b_i{\partial\over\partial x^a_i}\right)\cr\cr
D&=&x^+P^- +\sum_{i=1}^k\left(-{\partial\over\partial p_i^+}\, p_i^+
+\vec{x}_i \cdot {\partial\over\partial\vec{x}_i}\right)+{k(d-2)\over 2}\cr\cr
K^+&=&-{1\over 2}\sum_{i=1}^k\left(-2 x^+ {\partial\over\partial p_i^+}\, p_i^+
+\vec{x}_i\cdot\vec{x}_i \, p_i^+\right)+x^+D\cr\cr
K^-&=& \sum_{i=1}^k\Bigg({6-d\over 2}{\partial\over\partial p_i^+}
+p_i^+ {\partial^2\over\partial p_i^{+\,\,2}}
-\vec{x}_i\cdot {\partial\over\partial\vec{x}_i}{\partial\over\partial p_i^+} 
+{\vec{x}_i\cdot\vec{x}_i\over 4p_i^+}{\partial\over\partial\vec{x}_i}\cdot{\partial\over\partial\vec{x}_i}\Bigg)\cr\cr
K^a&=&-{1\over 2}\sum_{i=1}^k\left(-2 x^+ {\partial\over\partial x^a_i}{\partial\over\partial p_i^+}+\vec{x}_i\cdot\vec{x}_i {\partial\over\partial x^a_i}\right)\cr
&&+\sum_{i=1}^k x^a_i \left(-x^+ {1\over 2p_i^+}{\partial\over\partial\vec{x}_i}\cdot{\partial\over\partial\vec{x}_i}-{\partial\over\partial p_i^+}p_i^+ + \vec{x}_i\cdot{\partial\over\partial \vec{x}_i}+{d-2\over 2}\right)\label{CFTAlgebra}
\end{eqnarray}
For the bulk AdS$_{d+1}$ generators, once again, we rely on results obtained by Metsaev \cite{Metsaev:2003cu}. The formulas below are obtained by Fourier transforming (from $X^-$ to $P^+$) the generators given in Section 2 of \cite{Metsaev:2003cu}. The generators are
\bea
\vec{P}&=&{\partial\over\partial \vec{X}}\cr\cr
P^+&=&P^+\cr\cr
P^-&=&-{\partial_{\vec{X}}\cdot\partial_{\vec{X}}+\partial_Z^2\over 2P^+}+{1\over 2Z^2P^+}A\cr\cr
D&=&X^+ P^- -P^+\partial_{P^+}+\vec{X}\cdot\partial_{\vec{X}}+Z\partial_Z\cr\cr
J^{+-}&=&X^+P^-+P^+\partial_{P^+}+1\cr\cr
J^{+a}&=&X^+\partial_{X^a}-X^aP^+\cr\cr
J^{-a}&=&-\partial_{P^+}\partial_{X^a}-X^a P^-+M^{-a}\cr\cr
J^{ab}&=&X^a{\partial\over\partial X^b}-X^b{\partial\over\partial X^a}+M^{ab}\cr\cr
K^+&=&-{1\over 2}(\vec{X}\cdot\vec{X}+Z^2-2X^+\partial_{P^+})P^++X^+ D\cr\cr
K^a&=&-{1\over 2}(\vec{X}\cdot\vec{X}+Z^2-2X^+\partial_{P^+})\partial_{X^a}+X^a D+M^{aZ}Z+\sum_{b=1}^{d-2}M^{ab}X^b+M^{a-}X^+\cr\cr
K^-&=&-{1\over 2}(\vec{X}\cdot\vec{X}+Z^2-2X^+\partial_{P^+})P^--\partial_{P^+}D+{1\over P^+}\sum_{a=1}^{d-2}(X^a\partial_Z-Z\partial_{X^a})M^{aZ}\cr\cr
&&-\sum_{a=1}^{d-2}{X^a\over 2ZP^+}[M^{Za},A]+{1\over P^+}B\label{AdSAlgebra}
\eea
where $A$ is the AdS mass operator, $M^{AB}$ are the spin contributions to the Lorentz generators and $B$ is determined by the $M^{AB}$. The indices $A$ and $B$ run over $a$ and $Z$, i.e. they run over directions transverse to the light cone in the bulk. The form of these operators will be spelled out in complete detail below.

The bulk coordinates, relevant for the $k$-local collective field in $d$-dimensions, are determined by the requirement of bulk locality as explained in \cite{deMelloKoch:2024juz}. The bulk coordinates, expressed in terms of the coordinates of the collective field are
\bea
X^+&=&x^+\qquad\qquad P^+\,\,=\,\,\sum_{i=1}^k p_i^+\qquad\qquad Z\,\,=\,\,{\sqrt{\sum_{a=1}^{d-2}\sum_{i=1}^k p_i^+ (v_i^a)^2}\over (\sum_{l=1}^k p_i^+)^{3\over 2}}\cr\cr\cr
X^a&=&{\sum_{i=1}^k p_i^+ x_i^a\over \sum_{l=1}^k p_l^+}\qquad a=1,2,\cdots,d-2 \label{BulkCoords}
\eea
where 
\bea
v_i^a&=&\sum_{l=1}^k p_l^+ (x_i^a-x_l^a) \label{defvai}
\eea
We use $Z$ to denote the extra bulk holographic coordinate. The $d+1$ coordinates $X^+,P^+,X^a,Z$ parametrize the AdS$_{d+1}$ spacetime. Bulk locality also fixes the form of the spin generators \cite{deMelloKoch:2024juz}. The result is
\bea
M^{aZ}&=&{1\over Z(P^+)^2}\sum_{b=1}^{d-2}\sum_{i=1}^k v^b_i\left(v^a_j{\partial\over\partial x^b_j}-v^b_j{\partial\over\partial x^a_j}\right)\cr\cr
&+&{1\over 2Z(P^+)^2}\sum_{j,l=1}^k\left({p_l^+\over P^+}\sum_{b=1}^{d-2}(v^b_jv^b_j+v^b_lv^b_l){\partial\over\partial x^a_j}-2p_j^+p_l^+v^a_j{\partial\over\partial p^+_j}\right)
\eea
\bea
M_{ab}={1\over P^+}\sum_{l=1}^k\left(v_l^a{\partial\over\partial x_l^b}-v_l^b{\partial\over\partial x_l^a}\right)
\eea
The final spin generators $M^{-a}$ are determined using the relation \cite{Metsaev:1999ui}
\bea
M^{-a}&=&-M^{a-}\,\,=\,\,M^{aZ}{\partial_Z\over P^+}+\sum_{b=1}^{d-2}M^{ab}{\partial_b\over P^+}-{1\over 2ZP^+}\sum_{b=1}^{d-2}[M^{Zb},M^{ba}]
\eea
The AdS mass operator $A$ will be given below.

We now consider the matching of the so(2,$d$) generators acting on the collective field (denoted $L_{\rm CFT}$) with those acting on the bulk fields (denoted $L_{\rm AdS}$). Using the relation (\ref{relflds}) this amounts to requiring
\bea
L_{\rm AdS}=\mu(p_i^+,x_i) L_{\rm CFT}{1\over\mu(p_i^+,x_i)}\label{LeqL}
\eea
In practise we have performed the matching using mathematica. Matching the two sets of generators fixes
\bea
\mu(\vec{x}_i,p_i^+)&=&Z^{3-d\over 2}\,(P^+ Z)^{{k\over 2}(d-2)-1}\,(\prod_{i=1}^k p_i^+)^{4-d\over 2}\label{FormForMu}
\eea
We also obtain the following formula for the AdS mass operator
\bea
A&=&-{1\over 4(k-3)!}\sum_{i_1i_2\cdots i_{k-3}=1}^k\sum_{a=1}^{d-2}\kappa_{aa;i_1i_2\cdots i_{k-3}}\sum_{b=1}^{d-2}\kappa_{bb;i_1i_2\cdots i_{k-3}}\cr\cr
&&-{1\over 2(k-3)!}\sum_{a,b=1}^{d-2}\sum_{i_1i_2\cdots i_{k-3}=1}^k \left(\kappa_{ab;i_1i_2\cdots i_{k-3}}\kappa_{ab;i_1i_2\cdots i_{k-3}}-\kappa_{aa;i_1i_2\cdots i_{k-3}}\kappa_{bb;i_1i_2\cdots i_{k-3}}\right)\cr\cr\cr
&&-{1\over 2}\sum_{a,b=1}^{d-2}M_{ab}^2+{((k-1)d-2k+1)((k-1)d-2k-1)\over 4}
\label{AdSMassOp}
\eea
where
\bea
\kappa_{aa;i_1i_2\cdots i_{k-3}}&=&Z\sum_{i_1i_2\cdots i_{k-3}=1}^k\sum_{a=1}^{d-2}\epsilon_{i_1 i_2\cdots i_k}{\beta^a_{i_{k-2}}\sqrt{p^+_{i_{k-1}}}-\beta^a_{i_{k-1}}\sqrt{p^+_{i_{k-2}}}\over\sqrt{p^+_{i_k}}}{\partial\over\partial x^a_{i_k}}
\eea
\bea
\beta^a_i&=&{\sqrt{p_i^+}v^a_i\over Z (\sum_{l=1}^k p_l^+)^{3\over 2}}
\eea
and
\bea
\kappa_{ab;i_1i_2\cdots i_{k-3}}&=&{1\over 2}\left[ M_{ab},\kappa_{aa;i_1i_2\cdots i_{k-3}}\right]
\eea
Finally, the operator $B$ is given by
\bea
B&=&B^Z+\sum_{a=1}^{d-2}M^{Za}M^{Za}
\eea
where $B^Z$ is a component of the vector $B^I=(B^a,B^Z)$ which obeys
\bea
[B^I,B^J]+{1\over 2}\sum_{K,L}M^{IK}M^{KL}M^{LJ}-{1\over 2}\sum_{K,L}M^{JK}M^{KL}M^{LI}&&\cr\cr
-(C_{2,{\rm so(2,d)}}+\sum_{a=1}^{d-2}M^{aZ}M^{aZ}+2)M^{IJ}&=&0
\eea
In this last equation $K,L$ are summed over $a=1,2,\cdots,d-2$ and $Z$, and in this expression $C_{2,{\rm so(2,d)}}$ is the quadratic Casimir of the so(2,$d$) algebra (\ref{AdSAlgebra}). In practise we find that it is simpler to obtain $B$ by matching $K^-$ in (\ref{AdSAlgebra}) with $K^-$ in (\ref{CFTAlgebra}), rather than solving the above equation.

It is simple to verify that the relation between the fields given in (\ref{relflds}) with $\mu(\vec{x}_i,p_i^+)$ given in (\ref{FormForMu}), passes the simplest of all possible tests: it is dimensionally consistent. A scalar field $\phi(x^\mu)$ in $d$ dimensions has length dimension $L^{2-d\over 2}$. After taking a Fourier transform on one coordinate, the field $\phi(x^+,p^+,\vec{x})$ has length dimension $L^{4-d\over 2}$. Thus, the fluctuation $\eta_k(x^+,p_1^+,\vec{x}_1,\cdots,p_k^+,\vec{x}_k)$ of the $k$-local collective field in CFT$_d$ has length dimension $L^{k(4-d)\over 2}$. The same argument implies that the bulk AdS field $\Phi(X^+,P^+,\vec{X},Z,\{\alpha_i\})$ has length dimension $L^{3-d\over 2}$. The formula (\ref{relflds}) now implies that $\mu(\vec{x}_i,p_i^+)$ must have length dimension $L^{{3-d\over 2}+{k(d-4)\over 2}}$. It is simple to verify from (\ref{FormForMu}) that this is indeed the case.

We now have enough information to translate the bulk gravity equation of motion into the collective field theory. The bulk gravity equation of motion is \cite{Metsaev:2003cu}
\bea
\left(2P^+P^-+\sum_{a=1}^{d-2}\partial_{X^a}\partial_{X^a}+\partial_Z^2-{A\over Z^2}\right)\Phi=0
\eea
where the AdS mass operator $A$ is given by\footnote{We have used the explicit value of the Casimir of the SO(2,$d$) representation of the bulk gravity field. The representation is specified by a dimension $\Delta$ and the highest weight $(h_1, . . . h_\nu),$ $\nu=\lfloor{ d-1\over 2}\rfloor$ of a unitary representation of SO($d$). The bulk field is a totally symmetric and traceless field with $s$ indices and it is in a representation with dimension $\Delta$ and highest weight $(s,0,\cdots,0)$. The parameter $\Delta$ can be traded for a standard mass parameter using the formula $\Delta={d\over 2}+\sqrt{m^2+\left(s+{d-4\over 2}\right)^2}$. See \cite{Metsaev:2003cu} for more details.}
\bea
A&=&2B+{1\over 2}\sum_{a,b=1}^{d-2}M_{ab}^2+{(d+1)(d-1)\over 4}-\Delta(\Delta-d)-s(s+d-2)
\eea
Using the relation between the gravitational and collective fields (\ref{relflds}), the equation of motion can be written as
\bea
&&{1\over\mu(\vec{x}_i,p_i^+)}\Bigg(Z^2(2P^+P^-+\sum_{a=1}^{d-2}\partial_{X^a}\partial_{X^a}+\partial_Z^2)-\Big(2B+{1\over 2}\sum_{a,b=1}^{d-2}M_{ab}^2+{(d+1)(d-1)\over 4}\cr\cr
&&\quad -\Delta(\Delta-d)-s(s+d-2)\Big)\Bigg)\mu(\vec{x}_i,p_i^+) \,\eta_k(x^+,p_1^+,\vec{x}_1,\cdots,p_k^+,\vec{x}_k)\,\,=\,\,0
\eea
To proceed we note the following identity for the quadratic Casimir of the conformal field theory
\bea
C^{\rm CFT}_{2,{\rm SO(2,d)}}&=&{1\over\mu}\Bigg(Z^2(2P^+P^-+\sum_{a=1}^{d-2}\partial_{X^a}\partial_{X^a}+\partial_Z^2)-\Big(2B+{1\over 2}\sum_{a,b=1}^{d-2}M_{ab}^2\cr\cr
&&+{(d+1)(d-1)\over 4}\Big)\Bigg)\mu\label{casident}
\eea
where the quadratic Casimir of SO(2,d) ($M,N$ are summed over $-1,0,1,2,\cdots,d$) is given by
\bea
C^{\rm CFT}_{2,{\rm SO(2,d)}}&=&{1\over 2}L_{MN}L^{NM}\label{quadcas}
\eea
and we write $L_{MN}=-L_{NM}$ in terms of the conformal generators as ($\mu,\nu=0,1,\cdots,d-1$)
\bea
L_{\mu\nu}&=&J_{\mu\nu}\qquad L_{\mu\, -1}\,\,=\,\,{1\over 2}(P_\mu+K_\mu)\cr\cr
L_{-1\,d}&=&-D\qquad\quad L_{\mu 3}\,\,=\,\,{1\over 2}(P_\mu-K_\mu)
\eea
The indices $M$ and $N$ are raised and lowered with the metric $\eta={\rm diag}(-1,-1,1,1,\cdots,1)$ as usual. The identity (\ref{casident}) was established using mathematica. With the help of the identity (\ref{casident}) the equation of motion becomes
\bea
\Big(C^{\rm CFT}_{2,{\rm SO(2,d)}}-\Delta(\Delta-d)-s(s+d-2))\Big)\eta_k(x^+,p_1^+,\vec{x}_1,\cdots,p_k^+,\vec{x}_k)=0
\eea
This tells us that the bulk field is dual to the $(\Delta,s)$ component of $\eta_k(x^+,p_1^+,\vec{x}_1,\cdots,p_k^+,\vec{x}_k)$, exactly as we expect.

\section{Bulk coordinates}\label{Bcoords}

In the previous section we have obtained the formulas (\ref{BulkCoords}) which relate the bulk AdS$_{d+1}$ coordinates to a subset of the coordinates of the collective fields. In this section our goal is to obtain an interpretation for the remaining $(d-1)k -d$ coordinates. Our conclusion is that these are angles parametrizing the product S$^{k-1}\times$S$^{(d-2)(k-2)}\times$S$^{d-3}$ of spheres.

As a first step, it is useful to establish why these spheres appear in the geometry. First, light cone kinematics ensures that $p_i^+>0$. In addition, since
\bea
P^+=\sum_{i=1}^k p_i^+\label{FrstSphere}
\eea
it is natural to introduce $k-1$ angles defined as follows
\bea
p_1^+&=&P^+ \sin ^2(\psi_1)\cr\cr
p_2^+&=&P^+ \cos ^2(\psi_1) \sin ^2(\psi_2)\cr\cr
p_3^+&=&P^+ \cos ^2(\psi_1) \cos ^2(\psi_2) \sin ^2(\psi_3)\cr\cr
&\vdots&\cr\cr
p_k^+&=&P^+ \cos ^2(\psi_1) \cos ^2(\psi_2)\cdots \cos ^2(\psi_ {k-1})
\eea
These angles parametrize an S$^{k-1}$. Consequently, (\ref{FrstSphere}) is the origin of the first sphere.

To understand the origin of the remaining spheres, note that we can write the map from the bulk to the boundary for the spatial coordinates as ($a=1,2,\cdots,d-2$ and $i=1,2,\cdots k$)
\bea
x^a_i&=&X^a+\beta^a_i{\sqrt{P^+}Z\over\sqrt{p_i^+}}
\eea
where
\bea
\beta^a_i&=&{\sqrt{p_i^+}v^a_i\over Z (\sum_{l=1}^k p_l^+)^{3\over 2}}
\eea
and $v^a_i$ is defined in (\ref{defvai}). Using the definition of $Z$ it is easy to see that
\bea
\sum_{a=1}^{d-2}\sum_{i=1}^k (\beta^a_i)^2&=&1
\eea
while using the definition of $v_i^a$ we easily find
\bea
\sum_{i=1}^k\sqrt{p_i^+}\beta_i^a&=&0\qquad a=1,2,\cdots,d-2\label{const1}
\eea
Thus, the $k(d-2)$ quantities $\beta_i^a$ are components of a unit vector that obey $d-2$ constraints. This implies that there are a total of $k(d-2) -1-(d-2)$ independent angles contained in $\beta_i^a$. We will see below that they define an S$^{(k-2)(d-2)}\times$S$^{d-3}$ space. 

To parametrize the $\beta_i^a$ introduce the vector
\bea
(\hat{n}_{p^+})_j&=&\sin(\psi_j)\prod_{l=1}^{j-1}\cos(\psi_l)\quad j<k\cr\cr
&=&\prod_{l=1}^{k-1}\cos(\psi_l)\qquad\qquad j=k
\eea
which is parallel to the vector with components $v_i=\sqrt{p^+_i}$. Consequently, this vector plays an important role in enforcing the constraint (\ref{const1}). Following the construction of \cite{deMelloKoch:2024otg}, introduce the $k-1$ orthogonal unit vectors
\bea
\hat{n}_{(j)}&=&{1\over\prod_{i=1}^{j-1}\cos\psi_i}{\partial\over\partial\psi_j}\hat{n}_{p^+}
\eea
The collection of $k$ orthogonal vectors $\hat{n}_{p^+}$ and $\hat{n}_{(j)}$ with $j=1,2,\cdots,k-1$ provide an orthonormal basis for the space in which the sphere defined by (\ref{FrstSphere}) is embedded. The unit vector $\hat{n}_{p^+}$ is radially directed in this space. We also need to introduce a $d-2$ dimensional unit vector, defined by
\bea
N_a&=&\sin\phi_a\prod_{b=1}^{a-1}\cos\phi_b\qquad a<d-2\cr\cr
&=&\prod_{b=1}^{d-3}\cos\phi_b\qquad\qquad\quad a=d-2\label{defna}
\eea
where we have introduced $d-3$ angles $\phi_a$. Finally, introduce the angles $\alpha_i^a$ with $i=1,2,\cdots k-2$ and $a=1,2,\cdots,d-2$, and set $\alpha^a_{k-1}={\pi\over 2}$. With these definitions we can parametrize the $\beta_i^a$ as follows
\bea
\beta_i^a&=&N^aM_i\,\,=\,\,N^a\sum_{l=1}^{k-1}(\hat{n}_l)_i\sin(\alpha^a_l)\prod_{m=1}^{l-1}\cos(\alpha^a_m)
\eea
Here $N^a$ is a unit vector constructed using $d-3$ angles, defining an S$^{d-3}$, while $M_i$ is a unit vector constructed using $(k-2)(d-2)$ angles, defining an S$^{(k-2)(d-2)}$.

To summarize, the $k(d-1)$ coordinates $\{p_i^+,x^a_i\}$ with $a=1,2,\cdots,d-2$ and $i=1,2,\cdots,k$ are written in terms of the bulk coordinates $P^+,X^a,Z$ with $a=1,2,\cdots,d-2$, as well as the angles $\psi_j$ with $j=1,2,\cdots,k-1$, $\phi_a$ with $a=1,2,\cdots,d-3$ and $\alpha_i^a$ with $i=1,2,\cdots k-2$ and $a=1,2,\cdots,d-2$ as follows
\bea
p_i^+&=&P^+\sin(\psi_j)\prod_{l=1}^{i-1}\cos(\psi_l)\quad i<k\cr\cr
&=&P^+\prod_{l=1}^{k-1}\cos(\psi_l)\qquad\qquad i=k\cr\cr
x^a_i&=&X^a+N^a\sum_{l=1}^{k-1}(\hat{n}_l)_i\sin(\alpha^a_l)\prod_{m=1}^{l-1}\cos(\alpha^a_m){Z\over \sin(\psi_i)\prod_{r=1}^{i-1}\cos(\psi_r)}\qquad i<k\cr\cr
&=&X^a+N^a\sum_{l=1}^{k-1}(\hat{n}_l)_i\sin(\alpha^a_l)\prod_{m=1}^{l-1}\cos(\alpha^a_m){Z\over \prod_{r=1}^{k-1}\cos(\psi_r)}\qquad\qquad\quad i=k
\label{completeMap}
\eea
with $N_a$ defined in (\ref{defna}) and $\alpha^a_{k-1}={\pi\over 2}$. The $k$-local collective field itself defines a field on AdS$_{d+1}\times$S$^{k-1}\times$S$^{(k-2)(d-2)}\times$S$^{d-3}$. Notice that for $d=3$ this reduces to AdS$_4\times$S$^{k-1}\times$S$^{k-2}$ in agreement with \cite{deMelloKoch:2024otg}.

\section{Boundary Condition}\label{BC}

We have already explained in Section \ref{matchconf} how the bulk equation of motion is reproduced in the collective field theory. Each bulk field is in a definite SO(2,d) representation. The equation of motion, when translated to the collective field theory language, becomes the statement that we should extract the primary of the same representation from the $k$-local collective field. The fact that the equation of motion is recovered is not quite enough to claim that we have a correct bulk reconstruction. We must still show that the field has the correct boundary behaviour, i.e. that the GKPW rule is reproduced. Providing this demonstration is the goal of this section. 

To simplify the derivation of explicit formulas, we start with the specific case where $d=4$ and $k=2$
\bea
\sigma_2(x^+,p_1^+,x_1,y_1,p_2^+,x_2,y_2)&=&{\rm Tr}\left(\phi(x^+,p_1^+,x_1,y_1)\phi(x^+,p_2^+,x_2,y_2)\right)
\eea
This example is sufficiently involved that it nicely illustrates the features of the holographic map. In this scenario, the bilocal field lives in AdS$_5\times$S$^1\times$S$^1$. For the subsequent analysis, we need the detailed form of the holographic mapping as well as the behaviour of the bulk fields as the boundary is approached, i.e. as $Z\to 0$. First, let's explicitly outline the details of the holographic mapping. The collective field coordinates in terms of the bulk AdS coordinates ($X,Y,Z,P^+$) and an additional pair of angles $(\theta,\varphi$) are given by
\bea
x_1&=&X+\tan{\theta\over 2}\, Z\,\cos\varphi\qquad
x_2\,\,=\,\,X-\cot{\theta\over 2}\, Z\,\cos\varphi\cr\cr
y_1&=&Y+\tan{\theta\over 2}\, Z\,\sin\varphi\qquad
y_2\,\,=\,\,Y-\cot{\theta\over 2}\, Z\,\sin\varphi\cr\cr
p_1^+&=&P^+\cos^2{\theta\over 2}\qquad\qquad\qquad\qquad
p_2^+\,\,=\,\,P^+\sin^2{\theta\over 2}\label{d2k2map}
\eea
The inverse transformation is ($|\vec{x}_1-\vec{x}_2|\equiv \sqrt{(x_1-x_2)^2+(y_1-y_2)^2}$)
\bea
X&=&{p_1^+x_1+p_2^+x_2\over p_1^++p_2^+}\qquad\qquad\qquad
Y\,\,=\,\,{p_1^+y_1+p_2^+y_2\over p_1^++p_2^+}\cr\cr
P^+&=&p_1^++p_2^+\qquad\qquad\qquad\qquad
 Z\,\,=\,\, {\sqrt{p_1^+p_2^+}\over p_1^++p_2^+}|\vec{x}_1-\vec{x}_2|\cr\cr
\theta&=&2\arctan \sqrt{p_2^+\over p_1^+}\qquad\qquad
\tan\varphi\,\,=\,\,{y_1-y_2\over x_1-x_2}\label{id2k2map}
\eea
By matching translations, it is clear that the relation between the bulk and boundary momenta are
\bea
P^X&=&p_1^x+p_2^x\qquad\qquad P^Y\,\,=\,\,p_1^y+p_2^y\cr\cr
P^+&=&p_1^++p_2^+\qquad\qquad P^-\,\,=\,\,{(p_1^x)^2+(p_1^y)^2\over 2p_1^+}+{(p_2^x)^2+(p_2^y)^2\over 2p_2^+}
\eea
To find the momentum dual to $Z$ we use the chain rule
\bea
{\partial\over\partial Z}&=&{\partial x_1\over\partial Z}{\partial\over\partial x_1}+{\partial y_1\over\partial Z}{\partial\over\partial y_1}+{\partial x_2\over\partial Z}{\partial\over\partial x_2}+{\partial y_2\over\partial Z}{\partial\over\partial y_2}
\eea
together with (\ref{d2k2map}) to obtain
\bea
P_Z&=&{1\over 2}e^{-i\varphi}\left((p_1^x+i p_1^y)\tan{\theta\over 2}-(p_2^x+ip_2^y)\cot{\theta\over 2}\right)\cr\cr
&&\qquad +{1\over 2}e^{i\varphi}\left((p_1^x-ip_1^y)\tan{\theta\over 2}-(p_2^x-ip_2^y)\cot{\theta\over 2}\right)
\eea
In terms of
\bea
q^2&=&\left((p_1^x+ip_1^y)\tan{\theta\over 2}-(p_2^x + i p_2^y)\cot{\theta\over 2}\right)\left((p_1^x-ip_1^y)\tan{\theta\over 2}-(p_2^x-ip_2^y)\cot{\theta\over 2}\right)
\eea
and
\bea
\varphi_0&=&i\log\left({(p_1^x+ip_1^y)\tan{\theta\over 2}-(p_2^x + i p_2^y)\cot{\theta\over 2}\over q}\right)
\eea
we can write\footnote{We can also express the bulk momenta entirely in terms of the coordinates of the collective field
\bea
P^X&=&p^x_1+p^x_2\qquad P^Y\,\,=\,\,p^y_1+p^y_2\cr\cr
P^\varphi&=&{x_1-x_2\over |\vec{x}_1-\vec{x}_2|}(p_1^y-p_2^y)-{y_1-y_2\over |\vec{x}_1-\vec{x}_2|}(p_1^x-p_2^x)\cr\cr
P^Z&=&\sqrt{p_2^+\over p_1^+}\left({x_1-x_2\over |\vec{x}_1-\vec{x}_2|}p_1^x+{y_1-y_2\over |\vec{x}_1-\vec{x}_2|}p_1^y\right)-\sqrt{p_1^+\over p_2^+}\left({x_1-x_2\over |\vec{x}_1-\vec{x}_2|}p_2^x+{y_1-y_2\over |\vec{x}_1-\vec{x}_2|}p_2^y\right)
\eea
Note that we can easily verify that $dp_1^x dp_1^y dp_2^x dp_2^y\,=\,J dP^X dP^Y dP^Z dP^\varphi$ with $J={\sqrt{p_1^+p_2^+}\over 2(p_1^++p_2^+)}$.}  $P_Z=q\cos (\varphi+\varphi_0)$.

Now lets turn to behaviour of the bulk fields as $Z\to 0$. The bulk field obeys the wave equation
\bea
\left(2P^+i{\partial\over\partial x^+}+\partial_X^2+\partial_Y^2+\partial_Z^2-{A\over Z^2}\right)\Phi=0
\eea
where for symmetric massless fields the AdS mass operator simplifies to \cite{Metsaev:1999ui}
\bea
A&=&-{1\over 2}M^2_{ij}+{(d-3)(d-5)\over 4}\,\,=\,\,-M^2_{XY}-{1\over 4}
\eea
with $i,j$ running over $X,Y$ and in the second equality we have used $d=4$. Acting on a bulk field $\Phi_{(s)}$ that has a total of $s$ indices equal to $X$ or $Y$ this becomes
\bea
A\,\Phi_{(s)}&=&\left(s^2-{1\over 4}\right)\,\Phi_{(s)}\label{evalsofA}
\eea
Thus, the bulk wave equation becomes
\bea
\left(2P^+i{\partial\over\partial x^+}+\partial_X^2+\partial_Y^2+\partial_Z^2-{s^2-{1\over 4}\over Z^2}\right)\Phi_{(s)}&=&0
\eea
Now, introduce the operator $\hat{q}$ defined by \cite{Metsaev:2008fs}
\bea
\hat{q}^2&=&2P^+i{\partial\over\partial x^+}+\partial_X^2+\partial_Y^2\label{defq}
\eea
$\hat{q}$ commutes with both $Z$ and $\partial_Z$ so that is can be treated as a $c$-number. It is now simple to see that the normalizable solution to the bulk wave equation is given by
\bea
\Phi_{(s)}&=&\sqrt{Z}{J_s(\hat{q}Z)\over \hat{q}^s}f(x^+,P^+,X,Y)\label{solvebulkeom}
\eea
with $f(x^+,P^+,X,Y)$ set by the boundary condition obeyed by the field. Using the small $x$ behaviour $J_s(x)=x^s/(2^s s!)$, we find that for small $Z$ the field is given by
\bea
\Phi_{(s)}&=&Z^{s+{1\over 2}}f(x^+,P^+,X,Y)\left(1+O(Z)\right)
\eea
so that $f(x^+,P^+,X,Y)$ does indeed set the boundary behaviour. The GKPW rule would relate $f(x^+,P^+,X,Y)$ to an operator in the conformal field theory.

To determine the boundary behaviour $f(x^+,P^+,X,Y)$ predicted by collective field theory, we will now study the bulk field reconstructed from the collective fluctuation $\eta_2$. The construction of the bulk field in terms of the bilocal fluctuation is (use (\ref{relflds}) and (\ref{FormForMu}))
\bea
\Phi^{\rm bulk}(X^+,P^+,X,Y,Z,\{\theta,\varphi\})&=&P^+ \sqrt{Z}\,\,\eta_2(x^+,p_1^+,x_1,y_1,p_2^+,x_2,y_2)\label{recfld}
\eea
Write the fluctuation $\eta_2$ in terms of its Fourier transform and then rewrite the collective field's coordinates in terms of the bulk coordinates (using (\ref{d2k2map})) to obtain ($\{X^A\}\equiv\{X^+,P^+,X,Y,Z\}$)
\bea
\Phi^{\rm bulk}(\{X^A\},\{\theta,\varphi\})&=&\int_{-\infty}^\infty d^2 p_1 \int_{-\infty}^\infty d^2 p_2 \,\, e^{ip_1^x x_1+ip_2^x x_2+ip_1^y y_1+i p_2^y y_2}
P^+\sqrt{Z}\eta_2(x^{+}, p_{1}^{+}, \vec{p}_1, p_2^{+}, \vec{p}_2)\cr\cr
&=& \int_{-\infty}^\infty d^2 p_1 \int_{-\infty}^\infty d^2 p_2 \,\,e^{iX(p_1^x+p_2^x)+iY(p_1^y+p_2^y)}e^{i Zq\cos(\varphi+\varphi_0)}\cr\cr
&&\qquad\qquad\qquad\qquad
\times\, P^+\sqrt{Z}\,\eta_2(x^+, p_1^+,\vec{p}_1,p_2^+,\vec{p}_2)
\eea
Now, use the identity
\bea
\delta (\varphi)&=&{1\over 2\pi}\sum_{s=-\infty}^\infty e^{-is\varphi}
\eea
to obtain
\bea
\Phi^{\rm bulk}(\{X^A\},\{\theta,\varphi\})&=& \int_0^{2\pi}d\varphi'\,\delta(\varphi'-\varphi)\int_{-\infty}^\infty d^2 p_1 \int_{-\infty}^\infty d^2 p_2 \,\,e^{iX(p_1^x+p_2^x)+iY(p_1^y+p_2^y)}e^{i Zq\cos(\varphi+\varphi_0)}\cr\cr
&&\qquad\qquad\times\,P^+\sqrt{Z}\,\eta_2(x^+, p_1^+, p_1^x, p_1^y, p_2^+, p_2^x, p_2^y)\cr\cr
&=& \int_0^{2\pi}d\varphi'\,\delta(\varphi'-\varphi)\int_{-\infty}^\infty d^2 p_1 \int_{-\infty}^\infty d^2 p_2 \,\,e^{iX(p_1^x+p_2^x)+iY(p_1^y+p_2^y)}e^{i Zq\cos(\varphi'+\varphi_0)}\cr\cr
&&\qquad\qquad\times\,P^+\sqrt{Z}\,\eta_2(x^+, p_1^+, p_1^x, p_1^y, p_2^+, p_2^x, p_2^y)
\cr\cr
&=&{1\over 2\pi}\sum_{s=-\infty}^\infty \int_0^{2\pi}d\varphi'\,\int_{-\infty}^\infty d^2 p_1 \int_{-\infty}^\infty d^2 p_2 \,\, e^{-is(\varphi-\varphi')}e^{iX(p_1^x+p_2^x)+iY(p_1^y+p_2^y)}e^{i Zq\cos(\varphi'+\varphi_0)}\cr\cr
&&\qquad\qquad\times\,P^+\sqrt{Z}\,\eta_2(x^+, p_1^+, p_1^x, p_1^y, p_2^+, p_2^x, p_2^y)
\eea
The integral over $\varphi'$ can now be performed with the help of the Hansen-Bessel formula, which states
\bea
J_n(x)&=&{(-i)^n\over2\pi}\int_{-\pi}^\pi e^{ix\cos\varphi}e^{in\varphi}\, d\varphi
\eea
The result is
\bea
\Phi^{\rm bulk}(\{X^A\},\{\theta,\varphi\})&=&\sum_{s=-\infty}^\infty (-1)^s\,\int_{-\infty}^\infty d^2 p_1 \int_{-\infty}^\infty d^2 p_2 \,\, e^{-is\varphi}e^{iX(p_1^x+p_2^x)+iY(p_1^y+p_2^y)}\cr\cr
&&\qquad\times\,P^+\sqrt{Z}\,J_s(qZ)e^{-is\varphi_0}\eta_2(x^+, p_1^+, p_1^x, p_1^y, p_2^+, p_2^x, p_2^y)
\eea
Inserting the value for $\varphi_0$ and doing the momentum integrals, we obtain
\bea
\Phi^{\rm bulk}(\{X^A\},\{\theta,\varphi\})&=&\sum_{s=-\infty}^\infty e^{-is(\varphi+\pi)}\sqrt{Z}\,{J_s(\hat{q}Z)\over\hat{q}}f^{\rm rec}_s(x^+,p_1^+,p_2^+,X,Y)\label{nicebulky}
\eea
where
\bea
f^{\rm rec}_s(x^+,p_1^+,p_2^+,X,Y)&=& P^+\left(\sqrt{p_2^+\over p_1^+} \left(\partial_X^{(1)}+i\partial_Y^{(1)}\right)-\sqrt{p_1^+\over p_2^+}\left(\partial_X^{(2)}+i\partial_Y^{(2)}\right)\right)^s
\cr\cr
&&\times\,\eta_2(x^+, p_1^+,X,Y,p_2^+,X,Y)
\eea
The derivatives $\partial_X^{(i)}$ and $\partial_Y^{(i)}$ act on the $i^{\rm th}$ field in $\eta_2(x^+, p_1^+,X,Y,p_2^+,X,Y)$ with coordinates $x^+,p_i^+,X,Y$. $f^{{\rm rec},s}(x^+,P^+,X,Y)$ sets the boundary behaviour of the $e^{is\varphi}$ mode of the reconstructed bulk field. This operator should be related, by a GKPW rule, to operators in the dual conformal field theory. The fact that our reconstructed field (\ref{recfld}) takes exactly the form of a solution (\ref{solvebulkeom}) to the bulk equation of motion, as evidenced by (\ref{nicebulky}), is satisfying. It provides an independent confirmation of the conclusion reached in Section \ref{matchconf} that the collective field theory recovers the correct bulk equations of motion.

The collective field is defined on AdS$_5\times$S$^1\times$S$^1$. The first S$^1$ has angle $\varphi$ and the second has angle $\theta$. We have extracted the mode $e^{i\varphi}$ from the first S$^1$. Now we need to consider the second S$^1$. Recall that we have chosen a gauge that eliminates all $+$ and $-$ polarizations of the bulk higher spin fields. They are symmetric and traceless in their remaining indices, which run over $X,Y$ and $Z$. Thus, they must be in an irreducible representation of SO(3). The relevant spin generators have already been evaluated using bulk locality. The result is \cite{deMelloKoch:2024juz}
\bea
M^{XY}&=&{\partial\over\partial\varphi}
\eea
\bea
M^{XZ}&=&\cos\varphi{\partial\over\partial\theta}-\cot\theta\sin\varphi{\partial\over\partial\varphi}
\eea
\bea
M^{YZ}&=&\sin\varphi{\partial\over\partial\theta}+\cot\theta\cos\varphi{\partial\over\partial\varphi}
\eea
Since we want a mode in an SO(3) irreducible representation, we must require that
\bea
\Big(M^{XZ}M^{XZ}+M^{YZ}M^{YZ}+M^{XY}M^{XY}\Big)f_{s,l}(\theta,\varphi)&=&-l(l+1) f_{s,l}(\theta,\varphi)\label{eigenprob}
\eea
We have already determined the $\varphi$ dependence of our modes, which motivates the ansatz
\bea
f_{s,l}(\theta,\varphi)&=&e^{is\varphi} h_{s,l}(\theta)
\eea
Inserting this into (\ref{eigenprob}) we find that $h_{s,l}(\theta)$ must solve the equation
\bea
\left({d^2\over d\theta^2}+\cot\theta{d\over d\theta}+l(l+1)-{s^2\over\sin^2\theta}\right)h_{s,l}(\theta)&=&0
\eea
The solution is (this is in perfect agreement with the analysis of \cite{Mintun:2014gua} which uses different methods)
\bea
h_{s,l}(\theta)&=& c P^s_l(\cos\theta)
\eea
where $c$ is an arbitrary constant and $P^s_l(x)$ is the associated Legendre polynomial. Thus we learn that $0\le s\le l$. This is something we already know since at most $s=l$ of the indices on the bulk field can be $X,Y$ indices and at minimum none of the indices are $X,Y$ indices.

After we have reduced to the $e^{-is\varphi}$ mode we can now use the following orthogonality relation for the associated Legendre polynomials\footnote{The associated Legendre polynomials are not mutually orthogonal in general. However, some subsets are orthogonal, like those appearing in (\ref{orthogrel}). This orthogonality between a subset of the associated Legendre polynomials is all that is needed to extract the relevant bulk mode.}
\bea
\int_0^\pi P^m_k(\cos\theta)P^m_l(\cos\theta)\,\sin\theta\,d\theta&=&{2(l+m)!\over (2l+1)(l-m)!}\delta_{k,l}\label{orthogrel}
\eea
to obtain
\bea
f^{\rm rec}_{l,s}(x^+,p_1^+,p_2^+,X,Y)&=&\int_0^\pi d\theta\sin\theta P^s_l(\cos\theta)P^+\left(\sqrt{p_2^+\over p_1^+} \left(\partial_X^{(1)}+i\partial_Y^{(1)}\right)-\sqrt{p_1^+\over p_2^+}\left(\partial_X^{(2)}+i\partial_Y^{(2)}\right)\right)^s
\cr\cr
&&\times\,\eta_2(x^+, p_1^+,X,Y,p_2^+,X,Y)
\eea
Using the mapping (\ref{d2k2map}), it is simple to verify that
\bea
d\theta &=& {dp_1^+\over\sqrt{p_1^+p_2^+}}\qquad\qquad\sin\theta\,\,=\,\,{\sqrt{p_1^+p_2^+}\over p_1^++p_2^+}\qquad\qquad\cos\theta\,\,=\,\,{p_1^+-p_2^+\over p_1^++p_2^+}
\eea
so that the behaviour of the $l,s$ mode near the boundary is given by
\bea
f^{\rm rec}_{l,s}(x^+,P^+,X,Y)&=&\,\int_0^{P^+} dp_1^+ P^s_l\left({p_1^+-p_2^+\over p_1^++p_2^+}\right)\left({p_2^+\left(\partial_X^{(1)}+i\partial_Y^{(1)}\right)-p_1^+\left(\partial_X^{(2)}+i\partial_Y^{(2)}\right)\over \sqrt{p_1^+p_2^+}}\right)^s
\cr\cr
&&\times\,\eta_2(x^+, p_1^+,X,Y,p_2^+,X,Y)
\eea
Since the field components vanish when $p_i^+<0$ the integral over $p_1^+$ can be extended to run from $-\infty$ to $\infty$. Transforming to position space, we find
\bea
f^{\rm rec}_{l,s}(x^+,X^-,X,Y)&=&P^s_l\left({\partial_{X^-}^{(1)}-\partial_{X^-}^{(2)}\over \partial_{X^-}^{(1)}+\partial_{X^-}^{(2)}}\right)\left({\partial_{X^-}^{(2)} \left(\partial_X^{(1)}+i\partial_Y^{(1)}\right)-\partial_{X^-}^{(1)}\left(\partial_X^{(2)}+i\partial_Y^{(2)}\right)\over \sqrt{\partial_{X^-}^{(1)}\partial_{X^-}^{(2)}}}\right)^s
\cr\cr
&&\times\,\eta_2(x^+, X^-,X,Y,X^-,X,Y)\label{boundaryBehaviour}
\eea
The fact that square roots appear in the denominator of the second factor on the RHS is problematic and these square roots would make a matching to primary operators in the conformal field theory uncertain. Fortunately, all such factors cancel out in the complete RHS of (\ref{boundaryBehaviour}). To see this explicitly, note that the associated Legendre polynomial can be written as
\bea
P^s_l\left({\partial_{X^-}^{(1)}-\partial_{X^-}^{(2)}\over \partial_{X^-}^{(1)}+\partial_{X^-}^{(2)}}\right)&=&{(-1)^s\over 2^l l!}\left({\sqrt{\partial_{X^-}^{(1)}\partial_{X^-}^{(2)}}\over \partial_{X^-}^{(1)}+\partial_{X^-}^{(2)}}\right)^s {d^{l+s}\over dx^{l+s}}(x^2-1)^l\Bigg|_{x={\partial_{X^-}^{(1)}-\partial_{X^-}^{(2)}\over \partial_{X^-}^{(1)}+\partial_{X^-}^{(2)}}}
\eea
This determines the boundary behaviour of the bulk field reconstructed from the collective field. To complete the discussion of the GKPW rule for $k=2$, we should now interpret the conformal field theory operator defined by this boundary behaviour (\ref{boundaryBehaviour}).

Notice that due to the presence of inverse powers of $\partial_{X^-}^{(1)}+\partial_{X^-}^{(2)}$ on the right hand side above, the $Z\to 0$ limit of $\eta_2(x^{+}, p_{1}^{+}, x_1, y_1, p_2^{+}, x_2, y_2)$ is not a local operator. The GKPW dictionary is usually established in the de Donder gauge. The study \cite{Mintun:2014gua} has explored how this dictionary is modified by the transition from de Donder to light cone gauge. One finds exactly the structure given above: the GKPW dictionary becomes non-local and one must take $l$ derivatives of the bulk field, with respect to $X^-$, to recover a local conformal field theory operator. This is in manifest agreement with (\ref{boundaryBehaviour}). In Appendix \ref{App1} we show that $f^{\rm rec}_{l,s}(x^+,X^-,X,Y)$ is related to primary currents of the conformal field theory.

A parallel analysis can be conducted for the trilocal field. Repeating our discussion for the trilocal collective field is worthwhile because it resides on AdS$_4\times$S$^2\times$S$^2\times$S$^1$, making it the first example for which all three spheres participate. In addition to this motivation, the trilocal is an example of a massive spinning field, and the treatment of the massive case is more involved than the massless case. The coordinates of the collective field in terms of the coordinates of the gravity dual are given by
\bea
x_1&=&X+\sin\phi\Big(\cos\psi_1\sin\alpha^X\Big){Z\over\sin\psi_1}\cr\cr
x_2&=&X+\sin\phi\Big(-\sin\psi_1\sin\psi_2\sin\alpha^X+\cos\psi_2\cos\alpha^X\Big){Z\over \sin\psi_2\cos\psi_1}\cr\cr
x_3&=&X+\sin\phi\Big(-\sin\psi_1\cos\psi_2\sin\alpha^X-\sin\psi_2\cos\alpha^X\Big){Z\over\cos\psi_1\cos\psi_2}\cr\cr
y_1&=&Y+\cos\phi\Big(\cos\psi_1\sin\alpha^Y\Big){Z\over\sin\psi_1}\cr\cr
y_2&=&Y+\cos\phi\Big(-\sin\psi_1\sin\psi_2\sin\alpha^Y+\cos\psi_2\cos\alpha^Y\Big){Z\over \sin\psi_2\cos\psi_1}\cr\cr
y_3&=&Y+\cos\phi\Big(-\sin\psi_1\cos\psi_2\sin\alpha^Y-\sin\psi_2\cos\alpha^Y\Big){Z\over\cos\psi_1\cos\psi_2}\cr\cr
p_1^+&=&P^+\sin^2\psi_1\cr\cr
p_2^+&=&P^+\cos^2\psi_1\sin^2\psi_2\cr\cr
p_3^+&=&P^+\cos^2\psi_1\cos^2\psi_2
\eea
The angles $\psi_1$ and $\psi_2$ parametrize an S$^2$, the angles $\alpha^X$ and $\alpha^Y$ parametrize a second S$^2$ and the angle $\phi$ parametrizes an S$^1$. The inverse transformation is
\bea
X&=&{p_1^+x_1+p_2^+x_2+p_3^+x_3\over p_1^++p_2^++p_3^+}\qquad\qquad\qquad\qquad\qquad\qquad\,\,\,
Y\,\,=\,\,{p_1^+y_1+p_2^+y_2+p_3^+y_3\over p_1^++p_2^++p_3^+}\cr\cr
Z&=&\frac{\sqrt{\sum_{i>j=1}^3 p^+_i p^+_j \left((x_i-x_j)^2+(y_i-y_j)^2\right)}}{p^+_1+p^+_2+p^+_3}\qquad\qquad P^+\,\,=\,\,p_1^++p_2^++p_3^+\cr\cr
\psi_1&=&\tan^{-1}\sqrt{p_1^+\over p_2^++p_3^+}\qquad\qquad\qquad\qquad\qquad\qquad\qquad \psi_2\,\,=\,\,-\tan^{-1}\sqrt{p_2^+\over p_3^+}\cr\cr
\phi&=&\tan ^{-1}\left(\frac{\sqrt{p^+_1 (p^+_2 (x_1-x_2)+p^+_3 (x_1-x_3))^2+p^+_2 p^+_3 (x_2-x_3)^2 (p^+_1+p^+_2+p^+_3)}}{\sqrt{p^+_1 (p^+_2 (y_1-y_2)+p^+_3 (y_1-y_3))^2+p^+_2 p^+_3 (y_2-y_3)^2 (p^+_1+p^+_2+p^+_3)}}\right)\cr\cr
\alpha^X&=&-\tan^{-1}\left({\sqrt{p_1^+}(p_2^+(x_1-x_2)+p_3^+(x_1-x_3))\over\sqrt{p_2^+p_3^+(p_1^++p_2^++p_3^+)}(x_3-x_2)}\right)\cr\cr
\alpha^Y&=&-\tan^{-1}\left({\sqrt{p_1^+}(p_2^+(y_1-y_2)+p_3^+(y_1-y_3))\over\sqrt{p_2^+p_3^+(p_1^++p_2^++p_3^+)}(y_3-y_2)}\right)
\eea

To understand the behaviour of the bulk field as $Z\to 0$ we again need to study the bulk equation of motion. As a first step, it is again an instructive exercise to diagonalize the AdS mass operator. Just as for the case that $d=3$ studied in \cite{deMelloKoch:2024otg}, there is an SU(2) algebra structure underlying the mass operator. The mass operator can be written as
\bea
A&=&-L_z^2-2(L_+L_-+L_-L_+)+{3\over 4}\label{DiffAOp}
\eea
where the $L_m$s obey the algebra
\bea
[L_-,L_+]&=&-iL_z\qquad [L_z,L_\pm]\,\,=\,\,\pm iL_\pm
\eea
The explicit form of the $L_m$s is as follows
\bea
L_z&=&{1\over 2}(\kappa_{XX}-\kappa_{YY})\,\,=\,\,{\partial\over\partial\alpha^Y}-{\partial\over\partial\alpha^X}\cr\cr
L_+&=&{1\over 2}(\kappa_{XY}+iM_{XY})\,\,\equiv\,\,{1\over 2}(L_x+iL_y)\cr\cr
&=&e^{-i\alpha^X+i\alpha^Y}\Bigg(-{i\over 2}{\partial\over\partial\phi}
-{1\over 2\sin(2\phi)}\left({\partial\over\partial\alpha^X}+{\partial\over\partial\alpha^Y}\right)-{\cot(2\phi)\over 2}\left({\partial\over\partial\alpha^X}-{\partial\over\partial\alpha^Y}\right)\Bigg)\cr\cr
L_-&=&{1\over 2}(\kappa_{XY}-iM_{XY})\cr\cr
&=&e^{i\alpha^X-i\alpha^Y}\Bigg({i\over 2}{\partial\over\partial\phi}
-{1\over 2\sin(2\phi)}\left({\partial\over\partial\alpha^X}+{\partial\over\partial\alpha^Y}\right)-{\cot(2\phi)\over 2}\left({\partial\over\partial\alpha^X}-{\partial\over\partial\alpha^Y}\right)\Bigg)
\eea
The Wigner function
\bea
D^s_{m_1m_2}(\alpha^X,\alpha^Y,\phi)&=&e^{-im_1(\alpha^X+\alpha^Y)}d^s_{m_1m_2}\left(\cos(2\phi)\right)e^{im_2(\alpha^X-\alpha^Y)}
\eea
is a simultaneous eigenfunction of $L_z$ and the SU(2) Casimir, which again closely parallels results for $d=3$ \cite{deMelloKoch:2024otg}. It follows that the Wigner functions are a complete set of eigenfunctions of the AdS mass operator
\bea
A\, D^s_{m_1m_2}(\alpha^X,\alpha^Y,\phi)&=&\left((2s+1)^2-{1\over 4}\right)\, D^s_{m_1m_2}(\alpha^X,\alpha^Y,\phi)
\eea
Notice that just as we found for the massless case in (\ref{evalsofA}), the eigenvalues of the AdS mass operator are given by the square of an integer minus ${1\over 4}$. This ensures that the bulk wave function is again a Bessel function. Acting on a bulk field with dependence on $\alpha^X,\alpha^Y$ and $\phi$ fixed by the Wigner function
\bea
\Phi_{(s,m_1,m_2)}(\{X^A\},\{\alpha^A,\phi,\psi_i\})&=&F(\{X^A\},\{\psi_1,\psi_2\})D^s_{m_1m_2}(\alpha^X,\alpha^Y,\phi)
\eea
the bulk wave equation becomes
\bea
\left(2P^+i{\partial\over\partial x^+}+\partial_X^2+\partial_Y^2+\partial_Z^2-{(2s+1)^2-{1\over 4}\over Z^2}\right)\Phi_{(s,m_1,m_2)}(\{X^A\},\{\alpha^A,\phi,\psi_i\})&=&0\cr &&
\eea
Arguing exactly as we did above, it is simple to see that a normalizable solution to the bulk wave equation is given by
\bea
\Phi_{(s,m_1,m_2)}(\{X^A\},\{\alpha^A,\phi,\psi_i\})&=&\sqrt{Z}{J_{2s+1}(\hat{q}Z)\over \hat{q}^{2s+1}}g(x^+,P^+,X,Y,\psi_1,\psi_2)D^s_{m_1m_2}(\alpha^X,\alpha^Y,\phi)\cr
&&\label{solvesecondbulkeom}
\eea
with $g(x^+,P^+,X,Y,\psi_1,\psi_2)$ set by the boundary condition obeyed by the field. Using the small $x$ behaviour of the Bessel function, we find that for small $Z$ we have
\bea
\Phi_{(s,m_1,m_2)}(\{X^A\},\{\alpha^X,\alpha^Y,\phi\})&=&Z^{2s+{3\over 2}}g(x^+,P^+,X,Y,\psi_1,\psi_2)D^s_{m_1m_2}(\alpha^X,\alpha^Y,\phi)\left(1+O(Z)\right)
\nonumber
\eea
so that $g(x^+,P^+,X,Y,\psi_1,\psi_2)$ does indeed set the boundary behaviour.

We now explain how to construct the modes used to perform a harmonic expansion on S$^{k-1}\times$S$^{(d-2)(k-2)}\times$S$^{d-3}$. The construction proceeds by building the modes associated to a specific bulk field. We once again draw heavily on the work of Metsaev \cite{Metsaev:2003cu}, which examines massive spinning fields using a framework involving the SO(3) rotation group acting on the directions ($X,Y,Z$) transverse to the light cone in the bulk. The massive field\footnote{The massive fields are in an irreducible representation of SO(4). This SO(4) acts on all four spatial directions of the AdS$_5$ bulk.} of spin $s$ $\Phi^{I_1\cdots I_s}$ then decomposes as a direct sum of SO(3) representations with spins $s'=0,1,\cdots,s$. To ease the algebra of tensor indices, again following \cite{Metsaev:2003cu}, introduce the oscillators ($I,J=X,Y,Z$)
\bea
[\bar{a}^I,a^J]&=&\delta^{IJ}\qquad\qquad\bar{a}^I|0\rangle\,\,=\,\,0
\eea
The spin $s'$ field is then conveniently written as a ket vector as follows
\bea
|\phi_{s'}\rangle&\equiv& a^{I_1}\cdots a^{I_{s'}}\phi^{I_1\cdots I_{s'}}(x)|0\rangle
\eea
The fact that the field $\phi_{s'}$ is traceless and has spin $s'$ is written in terms of the ket as
\bea
\bar{a}^I\bar{a}^I|\phi_{s'}\rangle&=&0\qquad\qquad 
(a^I \bar{a}^I-s')|\phi_{s'}\rangle\,\,=\,\,0
\eea
In this language, the AdS mass operator is given by \cite{Metsaev:2003cu}
\bea
A&=&{1\over 2}\sum_{I,J}M^{IJ}M^{IJ}+{(d-1)(d+1)\over 4}+\Big( \Delta (\Delta-d)+s(s+2)\Big)\cr\cr
&&+2B^Z+\sum_{a=X,Y}M^{Za}M^{Za}
\eea
where the SO(3) rotation generators are
\bea
M^{IJ}&=&a^I\bar{a}^J- a^J\bar{a}^I
\eea
and the operator $B^Z$ acts as \cite{Metsaev:2003cu}
\bea
B^Z|\phi_{s'}\rangle&=&b(s,s'-1)A^Z_{s'-1}|\phi_{s'-1}\rangle+b(s,s')\bar{a}^Z|\phi_{s'+1}\rangle
\eea
with
\bea
b(s,s')&=&\sqrt{(s-s')(s+s'+2)(\Delta-s'-3)(\Delta-s'-1)\over 2s'+3}\qquad
A^Z_{s'-1}\,\,=\,\,a^Z-{\sum_J a^J a^J\bar{a}^Z\over 2s'+1}\nonumber
\eea
This completely determines the action of the AdS mass operator on the kets $|\phi_{s'}\rangle$.

It is now straight forward to determine the dependence of the mode on the angles to reproduce the behaviour a specific bulk state. First, in (\ref{DiffAOp}) we have established a representation of the AdS mass operator as a differential operator in the angles $\alpha^X$, $\alpha^Y$ and $\phi$. Demanding that Metsaev's action of the AdS mass operator on the bulk states is reproduced, fixes the dependence of the mode functions on these angles. Each state of the massive spinning field has a definite transformation under $M^{IJ}$ and is in a definite $SO(3)$ representation. The relevant SO(3) spin generators are easily determined. The result for $M^{XY}$ is given by
\bea
M^{XY}&=&-\cos(\alpha^X-\alpha^Y){\partial\over\partial\phi}+\sin(\alpha^X-\alpha^Y)\left(\cot(\phi){\partial\over\partial\alpha^X}+\tan(\phi){\partial\over\partial\alpha^Y}\right)
\eea
The results for $M^{XZ}$ and $M^{YZ}$ are given in Appendix \ref{UglyFormulas}. Both of these operators involve derivatives with respect to $\psi_1$ and $\psi_2$. The differential equations that ensure that the transformation of the mode functions match the transformation of the bulk states determines the dependence of the mode functions on the remaining angles $\psi_1$ and $\psi_2$.

An example will illustrate the procedure. The simplest example, that is non-trivial, corresponds to studying a massive field of spin 1. This representation has spin $s=1$ and dimension $\Delta=s+3=4$. The corresponding field is described using a pair of ket vectors
\bea
|\phi_1\rangle&=& a^X\phi_X|0\rangle+a^Y\phi_Y|0\rangle+a^Z\phi_Z|0\rangle\cr\cr
|\phi_0\rangle&=&\phi|0\rangle
\eea
Using Metsaev's representation, we easily find
\bea
A|\phi_1\rangle&=&{\Big(8\sqrt{3}\phi+11\phi_Z\Big)\over 4} a^Z|0\rangle +{15\over 4}\phi _X a^X|0\rangle +{15\over 4}\phi _Y a^Y|0\rangle\cr\cr
A|\phi_0\rangle&=&\left(2 \sqrt{3} \phi _Z+\frac{27 \phi }{4}\right)|0\rangle
\eea
Notice that $\phi_X$ and $\phi_Y$ are both eigenstates of the AdS mass operator, with eigenvalue given by $2^2-{1\over 4}$, again matching the massless case (\ref{evalsofA}). There are also linear combinations of $\phi$ and $\phi_Z$ that have eigenvalues given by $1^2-{1\over 4}$ and $3^2-{1\over 4}$, which is again of the form (\ref{evalsofA}). This again ensures that all the bulk wave functions are again Bessel functions. Since the Casimir of SO(3) evaluates to $l(l+1)$ for irreducible representation $l$, we have
\bea
\Big(M^{XZ}M^{XZ}+M^{YZ}M^{YZ}+M^{XY}M^{XY}\Big)|\phi_1\rangle&=&2|\phi_1\rangle\cr\cr
\Big(M^{XZ}M^{XZ}+M^{YZ}M^{YZ}+M^{XY}M^{XY}\Big)|\phi_0\rangle&=&0
\eea
In addition, it is simple to verify that
\bea
M^{XY}|\phi_0\rangle&=&0\qquad\qquad
M^{XY}|\phi_1\rangle\,\,=\,\,\phi _Y a^X|0\rangle -\phi _X a^Y|0\rangle\cr\cr
M^{XZ}|\phi_0\rangle&=&0\qquad\qquad
M^{XZ}|\phi_1\rangle\,\,=\,\,\phi _Z a^X|0\rangle -\phi _X a^Z|0\rangle\cr\cr
M^{YZ}|\phi_0\rangle&=&0\qquad\qquad
M^{YZ}|\phi_1\rangle\,\,=\,\,\phi _Z a^Y|0\rangle -\phi _Y a^Z|0\rangle
\eea
Using the realization of the AdS mass operator and the $M^{IJ}$ as differential operators, the above equations become the following differential equations
\bea
A\phi_Z&=&\frac{1}{4} \left(8\sqrt{3}\phi+11\phi_Z\right)\qquad A\phi\,\,=\,\,2\sqrt{3}\phi_Z+{27\over 4}\phi\cr\cr
A\phi_X&=&\left(2^2-\frac{1}{4}\right)\phi _X\qquad\qquad 
A\phi_Y\,\,=\,\,\left(2^2-\frac{1}{4}\right)\phi _Y\label{ModeDE1}
\eea
\bea
\Big(M^{XZ}M^{XZ}+M^{YZ}M^{YZ}+M^{XY}M^{XY}\Big)\phi_A&=&2\phi_A\qquad A=X,Y,Z\cr\cr
\Big(M^{XZ}M^{XZ}+M^{YZ}M^{YZ}+M^{XY}M^{XY}\Big)\phi&=&0\label{ModeDE2}
\eea
\bea
M^{XY}\phi&=&0\,\,=\,\, M^{XY}\phi_Z\qquad
M^{XY}\phi_X\,\,=\,\,\phi_Y\qquad\qquad M^{XY}\phi_Y\,\,=\,\,-\phi_X\cr\cr
M^{XZ}\phi&=&0\,\,=\,\, M^{XZ}\phi_Y\qquad
M^{XZ}\phi_X\,\,=\,\,\phi_Z\qquad\qquad M^{XZ}\phi_Z\,\,=\,\,-\phi_X\cr\cr
M^{YZ}\phi&=&0\,\,=\,\, M^{YZ}\phi_X\qquad
M^{YZ}\phi_Y\,\,=\,\,\phi_Z\qquad\qquad M^{YZ}\phi_Z\,\,=\,\,-\phi_Y\label{ModeDE3}
\eea
The solutions to (\ref{ModeDE1}), (\ref{ModeDE2}) and (\ref{ModeDE3}) define the mode functions $\{\phi,\phi_X,\phi_Y,\phi_Z\}$ needed to represent the massive bulk spin 1 field. Expanding the trilocal collective field in a complete set of modes, defined in this way, the coefficient of the above modes are the bulk fields dual to the $\Delta=4$ and $s=1$ primary multiplet in the conformal field theory.

From these two examples, we can recognize the general structure required to perform the mode decomposition of the collective field, thereby recovering the bulk fields of the dual gravity theory. First, the action of the AdS mass operator on bulk states is evaluated. Requiring that the mode functions reproduce this action implies a set of differential equations that fixes the dependence of the mode functions on the angles $\alpha_l^a$ and $\phi_a$. The eigenvalues of the AdS mass operator determine the fall off of the bulk fields as $Z\to 0$. Fields with a definite fall off correspond to spinning fields with a definite number of indices along the conformal field theory directions. The dependence on the angles $\psi_l$ is then determined by requiring that our mode functions transforms correctly under the action of the $M^{IJ}$ and that they are eigenfunctions of the SO($d$-1) Casimir
\bea
\sum_{a=1}^{d-2}M^{aZ}M^{aZ}+{1\over 2}\sum_{a,b=1}^{d-2}M^{ab}M^{ab}\label{SOdm1Cas}
\eea
This must be the case because our bulk field is a traceless and symmetric tensor with indices running over $X^a,Z$.

\section{Discussion and Conclusions}\label{conclusions}

We have outlined a comprehensive collective field theory description for the singlet sector of a free, massless matrix field in $d$ dimensions. Our results provide a complete framework for understanding large $N$ holography in the singlet sector of a free matrix model, extending the previous studies \cite{deMelloKoch:2024otg,deMelloKoch:2024ewt} which focused on $d=3$. Several new results were necessary to achieve this.

The mapping (\ref{relflds}) between the collective fields and the fields in the dual gravitational theory, defined on AdS$_{d+1}$ spacetime, involves a factor $\mu(\vec{x}_i, p_i^+)$. The form of this factor is novel and is specified in (\ref{FormForMu}). The $k$-local collective fields are now functions of $(d-1)k+1$ coordinates. The expressions for the bulk AdS$_{d+1}$ coordinates in terms of the collective field coordinates are provided in (\ref{BulkCoords}). These formulas are derived by applying the principle of bulk locality, as explained in \cite{deMelloKoch:2024juz}.

The remaining coordinates in the collective field are identified with angles, as described in (\ref{completeMap}). One of our key conclusions is that the coordinates of the collective field have a natural interpretation: the $k$-local collective field is defined on an AdS$_{d+1}\times$S$^{k-1}\times$S$^{(d-2)(k-2)}\times$S$^{d-3}$ spacetime.

In Section \ref{matchconf} we have derived an identity (see (\ref{casident})) which allowed us to conclude that the bulk equations of motion for the fields in SO(2,$d$) representation $(\Delta,s)$ become the statement in the collective field theory that the $(\Delta,s)$ primary should be extracted from the collective field. Consequently, the bulk equation of motion is accurately reflected in the collective field description. This conclusion was independently validated in Section \ref{BC} in formula (\ref{recfld}) which demonstrates that the correct form of the bulk solution is recovered from the fluctuation of the collective field. This analysis also confirms that the modes of a harmonic expansion on the S$^{k-1}\times$S$^{(d-2)(k-2)}\times$S$^{d-3}$ portion of the spacetime leads to the spinning bulk fields of the dual gravity theory. This analysis was performed for $d=4$ and $k=2,3$. This allowed us to consider the generic situation and the method outlined works for any $k$ and $d$.

We have outlined how the mode decomposition of the collective field, needed to recover the bulk fields of the dual gravity theory, is to be performed. The dependence of mode functions on the angles $\alpha_l^a$ and $\phi_a$ are fixed by reproducing the action of the AdS mass operator. The eigenvalues of the AdS mass operator control the fall off behaviour of the bulk fields as $Z\to 0$. The dependence of the mode functions on the angles $\psi_l$ is then determined by requiring that the mode function is an eigenfunction of the SO($d$-1) Casimir (\ref{SOdm1Cas}) and that it transforms correctly under the spin generators $M^{IJ}$. This completely fixes the procedure for how bulk fields are extracted from the collective field.

Our results, which give a constructive description of large $N$ holography for the singlet sector of a free matrix model, realize the proposal for holography first outlined for vector models in the prescient paper \cite{Das:2003vw} and realized in detail in \cite{deMelloKoch:2010wdf}. The fact that operators on the boundary $Z=0$ correspond to the limit in which all $k$ operators in the $k$-local collective field become coincident is closely related to the holographic nature of the collective field representation as explained in \cite{deMelloKoch:2023ylr}.

There are a number of directions in which this work can be extended. It is genuinely surprising that harmonic expansions on the spheres S$^{k-1}\times$S$^{(d-2)(k-2)}\times$S$^{d-3}$ organize the primary operators that are packaged into the $k$-local collective field. This construction, which also works for $d=3$ \cite{deMelloKoch:2024ewt}, suggests an interesting and novel structure underlying the conformal field theory. Understanding both the origins and further details of this construction is an extremely interesting problem.

Our analysis has focused on the free field theory. In the presence of interactions the collective field theory logic again dictates that the dynamics must be written in terms of invariant fields if the loop expansion parameter ${1\over N}$ is to be obtained. Understanding the interpretation of the extra coordinates (beyond those parametrizing the dual AdS$_{d+1}$ spacetime) is a fascinating problem. Solving this problem will shed light on how the primary operators in interacting conformal field theory are organized.

Another problem that deserves attention is the holographic mapping for equal time
collective fields. This would enable a study of finite temperature effects using the collective field theory description, which has the potential to shed light on holography for spacetimes
with horizons.

Finally, it would be interesting to probe ${1\over N}$ corrections to determine if the collective description can reproduce the non-linear interactions of the dual gravitational dynamics.

\begin{center} 
{\bf Acknowledgements}
\end{center}
RdMK is supported by a start up research fund of Huzhou University, a Zhejiang Province talent award and by a Changjiang Scholar award. HJRVZ is supported in part by the “Quantum Technologies for Sustainable Development” grant from the National Institute for Theoretical and Computational Sciences of South Africa (NITHECS).

\begin{appendix}

\section{Boundary behaviour and primary currents}\label{App1}

In Section \ref{BC} we have seen that the boundary ($Z\to 0$) behaviour of the bulk field reconstructed from the bilocal collective field, is given by
\bea
{\cal O}_{l,s}&=&
\left(\partial_{X^-}^{(1)}+\partial_{X^-}^{(2)}\right)^l P^s_l\left({\partial_{X^-}^{(1)}-\partial_{X^-}^{(2)}\over \partial_{X^-}^{(1)}+\partial_{X^-}^{(2)}}\right)\left({\partial_{X^-}^{(2)} \left(\partial_X^{(1)}+i\partial_Y^{(1)}\right)-\partial_{X^-}^{(1)}\left(\partial_X^{(2)}+i\partial_Y^{(2)}\right)\over \sqrt{\partial_{X^-}^{(1)}\partial_{X^-}^{(2)}}}\right)^s\cr\cr\cr
&&\qquad\qquad\qquad\times\,\eta_2(x^+, X^-,X,Y,X^-,X,Y)
\label{bndryBeh}
\eea
The derivative raised to the power of $l$ up front is needed to obtain a local operator. The need for these derivatives is a familiar feature of the light cone GKPW rule - see Section 4.2 of \cite{Mintun:2014gua}. The GKPW rule suggests that this operator should be related to primary operators of the conformal field theory. In this Appendix we will show that this is indeed the case, for a few chosen values of $l$ and $s$. Specifically we consider $l=2$ and any $s$, since this example is simple enough that we don't drown in details, while still complex enough that it is an instructive exercise. Our goal is simply to use the conformal algebra to identify the primary operators contained in the boundary limit of the reconstructed bulk field.

Our manipulations are all algebraic. We use the following properties of the free matrix scalar field
\bea
K^\mu\phi&=&0\qquad\qquad D\phi\,\,=\,\,\phi\qquad\qquad J^{\mu\nu}\phi\,\,=\,\,0
\eea
and the following pieces of the full conformal algebra
\bea
[K^\mu,P^\nu]&=&2\eta^{\mu\nu}D-2J^{\mu\nu}
\eea
\bea
[J^{\mu\nu},P^\alpha]&=&\eta^{\nu\alpha}P^\mu-\eta^{\mu\alpha}P^\nu
\eea
Using these results it is simple to prove that
\bea
K^+ (P^+)^k(P^x+iP^y)^s\phi&=&0\cr\cr
K^- (P^+)^k(P^x+iP^y)^s\phi &=&2k(k+s) (P^+)^{k-1}(P^x+iP^y)^s\phi\cr\cr
K^x (P^+)^k(P^x+iP^y)^s\phi &=&2s(s+k)(P^+)^k(P^x+iP^y)^{s-1}\phi\cr\cr
K^y (P^+)^k(P^x+iP^y)^s\phi &=&2is(s+K)(P^+)^k(P^x+iP^y)^{s-1}\phi
\eea
We can now study ${\cal O}_{l,s}$.

{\vskip 0.5cm}

\noindent
{\bf $l=2$ and $s=0$:} In this case the operator (\ref{bndryBeh}) when written in terms of components of the momentum becomes
\bea
{\cal O}_{2,0}=(P^+)^2\phi\,\phi-4 P^+\phi\, P^+\phi+\phi\, (P^+)^2\phi
\eea
We must take a trace to obtain a gauge invariant operator. It is simple to verify that
\bea
K^-{\cal O}_{2,0}\,\,=\,\,K^+{\cal O}_{2,0}\,\,=\,\,K^x{\cal O}_{2,0}\,\,=\,\,
K^y{\cal O}_{2,0}\,\,=\,\,0
\eea
In fact, we can prove more than this. It is simple to verify that
\bea
K^-{\cal O}_{l,0}\,\,=\,\,K^+{\cal O}_{l,0}\,\,=\,\,K^x{\cal O}_{l,0}\,\,
=\,\,K^y{\cal O}_{l,0}\,\,=\,\,0
\eea
using the formulas given above. The fact that these operators are primary is in good agreement with \cite{Mintun:2014gua,deMelloKoch:2014vnt}.

{\vskip 0.5cm}

\noindent
{\bf $l=2$ and $s=1$:} In this case the operator (\ref{bndryBeh}) when written in terms of components of the momentum becomes
\bea
{\cal O}_{2,1}&=&(P^+)^2\phi\, (P^x+iP^y)\phi- P^+ (P^x+iP^y)\phi\, P^+\phi -P^+\phi\, P^+ (P^x+i P^y)\phi\cr\cr
&&+ (P^x+iP^y)\phi\, (P^+)^2\phi 
\eea
Again, we must take a trace to obtain a gauge invariant operator.

It is simple to verify that
\bea
K^+{\cal O}_{2,1}&=&0\qquad\qquad\qquad K^-{\cal O}_{2,1}\,\,=\,\,{\cal O}_A\cr\cr
K^x{\cal O}_{2,1}&=&2{\cal O}_{2,0}\qquad\qquad K^y{\cal O}_{2,1}\,\,=\,\, 2i{\cal O}_{2,0}
\eea
where
\bea
{\cal O}_A&=&4P^+\phi\,(P^x+iP^y)\phi-2P^+(P^x+iP^y)\phi\,\phi-2\phi\,P^+(P^x+iP^y)\phi\cr\cr
&&+4(P^x+iP^y)\phi\, P^+\phi
\eea
Above we have proved that ${\cal O}_{2,0}$ is primary. It is simple to show that
\bea
K^+{\cal O}_A&=&K^-{\cal O}_A\,\,=\,\,K^x{\cal O}_A\,\,=\,\,K^y{\cal O}_A\,\,=\,\,0
\eea
i.e. ${\cal O}_A$ is also primary. Thus, ${\cal O}_{2,1}$ is a linear combination of two level 1 descendants. This reproduces the light cone GKPW rule of \cite{Mintun:2014gua}.

{\vskip 0.5cm}

\noindent
{\bf $l=2$ and $s=2$:} The operator (\ref{bndryBeh}) when written in terms of components of the momentum is
\bea
{\cal O}_{2,2}&=&(P^+)^2\phi\,(P^x+i P^y)^2\phi -2 P^+(P^x+i P^y)\phi\, P^+(P^x+i P^y)\phi\cr\cr
&&+ (P^x+i P^y)^2\phi\,(P^+)^2\phi
\eea
It is simple to verify that $K^+{\cal O}_{2,2}=0$ and
\bea
K^-{\cal O}_{2,2}&=&8P^+\phi\,(P^x+iP^y)^2\phi-8(P^x+iP^y)\phi\,P^+(P^x+iP^y)\phi \cr\cr
&&-8P^+(P^x+iP^y)\phi\,(P^x+iP^y)\phi+8\,(P^x+iP^y)^2\phi\,P^+\phi\cr\cr
&&\equiv\,\,{\cal O}_B
\eea
\bea
K^x{\cal O}_{2,2}&=&8(P^+)^2\phi\,(P^x+iP^y)\phi-8P^+\phi\,P^+(P^x+iP^y)\phi \cr\cr
&&-8P^+(P^x+iP^y)\phi\,P^+\phi+8\,(P^x+iP^y)\phi\,(P^+)^2\phi\cr\cr
&&\equiv\,\,{\cal O}_C
\eea
and $K^y{\cal O}_{2,2}=i{\cal O}_C$. It is now easy to prove that $K^+{\cal O}_B=0$,
\bea
K^-{\cal O}_B&=&16\phi\,(P^x+iP^y)^2\phi-64 (P^x+iP^y)\phi\,(P^x+iP^y)\phi\cr\cr
&&+16(P^x+iP^y)^2\phi\,\phi\cr\cr
&\equiv& {\cal O}_D
\eea
\bea
K^x{\cal O}_B&=&32 P^+\phi\,(P^x+iP^y)\phi+32(P^x+iP^y)\phi\, P^+\phi-16 P^+(P^x+iP^y)\phi\,\phi\cr\cr
&&-16\phi\, P^+(P^x+iP^y)\phi\cr\cr
&\equiv&{\cal O}_E
\eea
$K^y=i{\cal O}_E$, $K^+{\cal O}_C=0$, $K^-{\cal O}_C={\cal O}_E$, $K^x{\cal O}_C=16{\cal O}_{2,0}$, $K^y{\cal O}_C=16i{\cal O}_{2,0}$, as well as $K^+{\cal O}_D=0$, $K^-{\cal O}_D=0$, $K^x{\cal O}_D=0$, $K^y{\cal O}_D=0$, $K^+{\cal O}_E=0$, $K^-{\cal O}_E=0$, $K^x{\cal O}_E=0$ and $K^y{\cal O}_E=0$. This proves that ${\cal O}_{2,2}$ is a linear combination of level 2 descendants.

Based on the examples we have studied here, it is natural to conjecture that the operator ${\cal O}_{l,s}$ will be a linear combination of level $s$ descendants.

\section{Operators in terms of angles}\label{UglyFormulas}

In this section we collect some formulas that were useful for the analysis performed in Section \ref{BC}.

\bea
\kappa_{XX}&=&-2{\partial\over\partial\alpha^X}\qquad\qquad\qquad
\kappa_{YY}\,\,=\,\,-2{\partial\over\partial\alpha^Y}
\eea
\bea
\kappa_{XY}&=&-\sin(\alpha^X-\alpha^Y){\partial\over\partial\phi}-\cos(\alpha^X-\alpha^Y)\left(\cot(\phi){\partial\over\partial\alpha^X}+\tan(\phi){\partial\over\partial\alpha^Y}\right)
\eea

\bea
M^{XZ}&=&\frac{\cos(\phi)}{8}\left(\cos (\alpha^X) \Big(4 \sin (2 \alpha^Y) \tan (\psi_1)-4 {\cos ^2(\alpha^Y)\cos (2 \psi_2)\sec (\psi_2)\over \cos(\psi_1)\sin(\psi_2)}\Big)\right.\cr\cr
&&\left.\qquad\qquad\qquad\qquad+{\sin (\alpha^X)\left((\cos (2 \alpha^Y)-3) \cos (2 \psi_1)+2 \cos ^2(\alpha^Y)\right)\over \sin(\psi_1)\cos(\psi_1)}\right){\partial\over\partial\phi}\cr\cr
&&+\frac{1}{16} \left(8 \sin (\alpha^X) \cos (\phi ) \cot (\phi ) \left({\cos ^2(\alpha^Y) \cos(2\psi_2)\over\cos(\psi_1)\sin(\psi_2)\cos(\psi_2)}-\sin (2 \alpha^Y) \tan (\psi_1)\right)\right.\cr\cr
&&\left.\qquad\qquad+\cos(\alpha^X) \left({8\tan(\psi_1)\over\sin(\phi)}-8 \sin ^2(\alpha^Y) \cot (\psi_1) \cos (\phi ) \cot (\phi )\right)\right){\partial\over\partial\alpha^X}\cr\cr
&&+\frac{1}{2} \sin (\alpha^Y) \tan (\psi_1) \sin (\phi) \Big(\cos (\alpha^X) (2 \sin (\alpha^Y)-2 \cos (\alpha^Y) \csc (\psi_1) \cot (2 \psi_2))\cr\cr
&&\qquad\qquad\qquad+\sin (\alpha^X) \cos (\alpha^Y) \cot ^2(\psi_1)\Big){\partial\over\partial\alpha^Y}\cr\cr
&&-\frac{1}{2} \sin (\alpha^X) \sin (\phi){\partial\over\partial\psi_1}
-\frac{1}{2} \cos (\alpha^X) \sec (\psi_1) \sin (\phi){\partial\over\partial\psi_2}
\eea

\bea
M^{YZ}&=&-\frac{1}{16} \sin(\phi )\Big(-8{\cos^2(\alpha^X)\cos(\alpha^Y)\cos (2\psi_2)  \over\cos(\psi_1)\sin(\psi_2)\cos(\psi_2)}+8\sin(2\alpha^X)\cos (\alpha^Y)\tan (\psi_1)\cr\cr
&&+{\sin (\alpha^Y)(\cos (2\alpha^X-2\psi_1)+\cos (2\alpha^X+2\psi_1)\over\sin(\psi_1)\cos(\psi_1)}+2 \cos (2 \alpha^X)-6\cos (2\psi_1)+2)\Big)\frac{\partial}{\partial \phi }\cr\cr
&&+\frac{1}{2}\sin(\alpha^X)\tan (\psi_1)\cos (\phi )\Bigg(\cos(\alpha^X) \Big(\sin (\alpha^Y) \cot ^2(\psi_1)-2{\cos(\alpha^Y)\over \sin(\psi_1)}\cot (2\psi_2)\Big)\cr\cr
&&\qquad\qquad\qquad\qquad\qquad +2 \sin (\alpha^X) \cos (\alpha^Y)\Bigg)\frac{\partial}{\partial\alpha^X}\cr\cr
&&+\frac{1}{4} \Bigg(2 \cos (\alpha^X) \tan (\psi_1) \sin (\phi ) \tan (\phi ) \Big(\cos (\alpha^X) \Big({\sin(\alpha^Y)\cos(2\psi_2)\over\sin(\psi_1)\sin(\psi_2)\cos(\psi_2)}+\cos (\alpha^Y)\Big)\cr\cr
&&-2\sin(\alpha^X)\sin(\alpha^Y)\Big)+{\cos(\alpha^Y)\over\sin(\psi_1)\cos(\psi_1)\cos(\phi)}\Big(\cos (2 \psi_1) \big(\cos (2 \alpha^X) \sin ^2(\phi )-1\big)\cr\cr
&&\qquad\qquad\qquad\qquad\qquad
+\cos ^2(\phi )\Big)\Bigg)\frac{\partial}{\partial \alpha^Y}\cr\cr
&&-\frac{1}{2} \cos (\alpha^Y) \sec (\psi_1) \cos (\phi ) \frac{\partial}{\partial \psi_2}
-\frac{1}{2} \sin (\alpha^Y) \cos (\phi ) \frac{\partial}{\partial \psi_1}
\eea

\end{appendix}

\end{document}